\documentclass[11pt]{imsart}

\usepackage{graphicx}
\usepackage{amssymb}
\usepackage{amsmath}
\usepackage{amsthm}
\usepackage{mathtools}
\usepackage{caption}
\usepackage{subcaption}
\usepackage[ruled,vlined]{algorithm2e}
\usepackage{float}
\newtheorem{theorem}{Theorem}
\newtheorem{lemma}[theorem]{Lemma}
\newtheorem{proposition}[theorem]{Proposition}
\newtheorem{corollary}[theorem]{Corollary}
\newtheorem{definition}[theorem]{Definition}

\let\oldproofname=\proofname
\renewcommand{\proofname}{\rm\bf{\oldproofname}}

\DeclarePairedDelimiter{\floor}{\lfloor}{\rfloor}

\startlocaldefs
\endlocaldefs

\graphicspath{ {images/} }

\begin{document}

\begin{frontmatter}

% "Title of the paper"
\title{Image Segmentation, Compression and Reconstruction from Edge Distribution Estimation with Random Field and Random Cluster Theories}
\runtitle{Edge Distribution Estimation}

\author{\fnms{Robert A.} \snm{Murphy, Ph.D.}\ead[label=e1]{robert.a.murphy@wustl.edu}}
\address{\printead{e1}}

\runauthor{Murphy}

\begin{abstract}
Random field and random cluster theory are used to describe certain mathematical results concerning the probability distribution of image pixel intensities characterized as generic $2D$ integer arrays.  The size of the smallest bounded region within an image is estimated for segmenting an image, from which, the equilibrium distribution of intensities can be recovered.  From the estimated bounded regions, properties of the sub-optimal and equilibrium distributions of intensities are derived, which leads to an image compression methodology whereby only slightly more than half of all pixels are required for a worst-case reconstruction of the original image.  A custom deep belief network and heuristic allows for the unsupervised segmentation, detection and localization of objects in an image.  An example illustrates the mathematical results.
\end{abstract}

%\begin{keyword}[class=MSC]
%\kwd[Primary ]{05C80}
%\kwd{KGK}
%\kwd{05C40}
%\kwd[; secondary ]{KG57}
%\kwd{82B41}
%\end{keyword}

\begin{keyword}
\kwd{markov property}
\kwd{random field}
\kwd{random cluster}
\kwd{deep belief network}
\kwd{segmentation}
\kwd{compression}
\kwd{reconstruction}
\end{keyword}

\end{frontmatter}

\newpage

\tableofcontents

\newpage

% AOS,AOAS: If there are supplements please fill:
%\begin{supplement}[id=suppA]
%  \sname{Supplement A}
%  \stitle{Title}
%  \slink[doi]{10.1214/00-AOASXXXXSUPP}
%  \sdatatype{.pdf}" 
%  \sdescription{Some text}
%\end{supplement}

\section{Introduction and Related Work}
\label{intro}

\begin{figure}[H]
\centering
\includegraphics[scale=0.75]{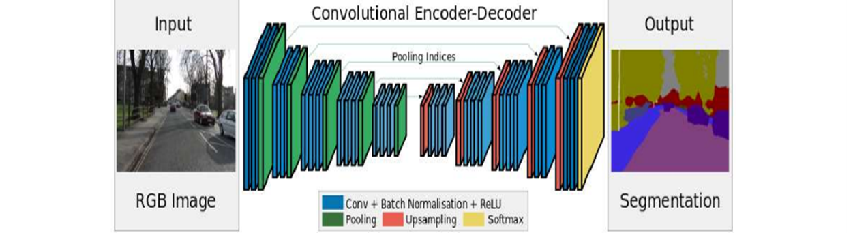}
\caption{Convolutional neural network with an input layer, (de) convolution and pooling/upsampling layers, compressed representational layers and an output layer, from $\cite{Holder}$}
\end{figure}

\subsection{Motivation}
\label{motisec}

Suppose an image in the $2D$ plane is uniformly partitioned into individual small square \textbf{pixels} of a predetermined size (necessarily defining some $R \times C$ resolution) to form a $2D$ array in $\mathbb{Z}^{2}$.  Assigned to each pixel are integer coordinates $(r,c) \in \mathbb{Z}^{2}$, where $2 \le r \le R$ and $2 \le c \le C$, and an integer value $i_{(r,c)} \in \mathbb{N}$ represents its color \textbf{intensity}.  In the $2D$ array, let two pixels be \textbf{neighbors} if their Euclidean distance is exactly $1$, virtually constructing an associated \textbf{edge} between the two neighbors.  Let an edge be \textbf{open}, if neighboring intensities are equivalent, within some error margin (yet to be determined heuristically) and \textbf{closed} otherwise.\\

From $\cite{Girshick,Girshick2,Gkioxari,He,Lin,Liu,Redmon,Ren}$, image object detection, segmentation and classification typically follows the supervised learning paradigm of:

\begin{itemize}
\item Abstraction via convolution and down-samplings.
\item Regional alignment of abstractions with their original image counterparts.
\item Localization (bounding box, mask or mesh).
\item Classification.
\item Approximation via de-convolution and up-samplings.
\end{itemize}

In particular, convolution layers operate on the intensities within small, overlapping sub-regions of an image.  Within a sub-region, convolutions are an average of the function of its center, weighted by a function of the surrounding values.  Convolution removes corners formed by intersecting line segments in the sub-region, delineated by sharp changes in intensity.  Sharp changes in intensity are then replaced by a more fluid transition, resulting in smooth curves that were once pointed intersection of line segments.

For reasons that will become clearer in latter sections, linear functions are applied to the centers of sub-regions, while Gaussian functions are applied to its surrounding values.  A Gaussian field of intensities is the resulting image obtained after convolutions are applied to all overlapping sub-regions.  By design, this field is log-multi-linearly separable, allowing for the log-linear separation of neighboring intensities within a sub-region from all other intensities in other sub-regions, providing a method for feature extraction.

Other (classical) image feature extraction heuristics highlight the boundary between objects of interest, when delineation of objects within an image is required by the problem space, such as human-animal clustering of pixels in a \textit{semantic segmentation}.  One such feature extraction heuristic, termed \textit{histogram of oriented gradients (HOG) descriptors} $\cite{Alhakeem,Alhindi,Huang,Kachouane,Kitayama,Soler}$, is a technique that locates and orients vectors in the direction of greatest change between intensities of neighboring pixels in an image, resulting in the highlighting of a \textit{negative space} around objects of interest.  Then, feature extraction via convolutions in a deep architecture is an open-edge detector by smoothing intersections within bounded sub-regions.  And, feature extractors of the same vein as the HOG descriptor heuristic are closed-edge detectors by highlighting the same intersections, yet leaving them unchanged.

Restricting our attention to a semantic segmentation, we can describe synergies between deep architectures and edge detectors, since every open-edge detector is, by default, also a closed-edge detector (and vice-versa), when you consider the negative space associated to all open (or closed) edges.  Such a synergistic deep architecture removes the extra burdens required during construction of HOG descriptors and the like, which includes normalizing transformations applied to pixel intensities, as well as selection and application of a clustering technique for vector orientations in localized regions, etc.  However, a commonality between segmentation and the HOG heuristic is pixel intensity (edge) density estimation associated to an unknown probability distribution $\mathbf{P}$ from which a \textbf{point process} $\mathcal{P}$ samples to construct an image.

\subsection{Metropolis Algorithm}
\label{metsec}

Let $\mathbf{E}$ be the set of all edges across the bounded region $\mathcal{B}$ defined by the set of pixels in an image in a predefined integer coordinate system.  Let $\mathbf{S}=\{-1,+1\}$ be the set of possible states for open and closed edges in $\mathbf{E}$, where $open=+1$.  Let $\mathbf{W}$ be the set of binary words over $\mathbf{E}$ and let $H(\mathbf{x},\beta)$ be a multi-linear function for fixed $\beta \in \mathbb{R}$ and $\mathbf{x} \in \mathbf{W}$.  Then, from $\cite{Guyon}$,

\begin{definition}
A \textbf{random field} is a normalized exponential distribution $\mathbf{P}$, defined as a function of some multi-linear function, $H(\mathbf{x},\beta)$.
\end{definition}

As in $\cite{Guyon}$, we say that a random field $\mathbf{P}$ is \textbf{separable}, if the covariance of $\mathbf{P}$ can be written as the product of the variances of the marginals of $\mathbf{P}$.  This definition of $\mathbf{P}$ necessarily requires that the image pixel intensities be modeled as a (separable) product of log-multi-linear processes, since the edge states are log-multi-linearly separable.

From $\cite{Geman}$, $\mathbf{P}$ can be estimated as the global distribution of a collection of single-edge conditional distributions of $\mathbf{P}$ using a modified Metropolis algorithm applied to randomly selected edges in $\mathbf{E}$.  A scale parameter $\beta$ in the log-multi-linear function is continuously adjusted according to a particular annealing schedule and each randomly selected edge state is changed to the opposite state in $\mathbf{S}$.  After a state change is applied, if the computed value of the conditional of $\mathbf{P}$ increases, which indicates decreasing energy (entropy), then the changed edge state is maintained.  Otherwise, the changed edge state is kept with a certain (predefined) uniform probability, $p \in (0,1)$.

A greedy algorithm of this sort allows us to check all local maxima while searching for the global, maximizing, equilibrium distribution of edge states for the current image, without getting stuck in local maxima.  Any changed edge states resulting in a lower value of a conditional of $\mathbf{P}$ are reverted to the previous edge state with non-zero probability, $1-p$.  In $\cite{Geman}$, it is shown that this algorithm necessarily leads to a global, energy minimizing, posterior probability distribution $\mathbf{P}$ for the current image.  From $\cite[Thm.\ (8.1)]{Grimmett}$, we know that $\mathbf{P}$ is unique.

Checking all local maxima using a modified Metropolis algorithm $\cite{Metropolis}$ can be very resource and time consuming.  Each adjustment of the annealing schedule results in a change to the scale parameter $\beta$ in the log-multi-linear energy function.  A series of samples of states are obtained from the collection of single-edge conditional distributions of $\mathbf{P}$ to have their respective state values changed to the other state value in $\mathbf{S}$.  After computation of the associated conditional of $\mathbf{P}$ and a decision is made whether to keep the state change, another conditional of $\mathbf{P}$ is sampled.  The process continues at the current value of the scale parameter $\beta$ until there is a relatively small difference in changes to the output value from the conditionals of $\mathbf{P}$ at the last two sampled edges, indicating (local) equilibrium state has been attained.

It is easy to see that if the number of stopping times $N$ in the schedule for the scale parameter $\beta$ is large and the number of samples $m$ from the collection of conditionals of $\mathbf{P}$ is large at each individual stopping time in the schedule, then convergence to the global, energy minimizing, posterior probability distribution $\mathbf{P}$ will be slow.  In fact, convergence times will grow exponentially as $\mathcal{O}(N^{m})$.

Take note of the main goal of the algorithm.  At each stopping time of the schedule, we perform a series of random selections from the collection of single-edge, conditional distributions of $\mathbf{P}$ and change the state at each edge in the sample.  If the edge’s state is changed from $closed=-1$ to $open=+1$, then this change is tantamount to requiring that the pixel intensities be equivalent at each site comprising the edge in question.  Likewise, changing an edge’s state from $open=+1$ to $closed=-1$ has the opposite effect on its associated pixel intensities, breaking symmetry.  Thus, suppose we can devise a deep learner with the capability of predicting local pixel intensities in small, uniformly sized regions of an image.  If those regions overlap so that we have a way to open and close the same edges within the same region, then we can possibly duplicate the modified Metropolis algorithm with more efficient processing times using the deep learner.

\subsection{Lattice Model}
\label{lm}

In $\cite{Murphy}$, it is proven that a closed-form, connectivity radius $\mathcal{R}$ can be constructed for high dimensional data such that all edges are open and connected within a fixed, bounded region, if the Euclidean distance between the data points is greater than $\mathcal{R}$.  Otherwise, disjoint sets of connected edges form when the distance between data points is less than or equal to $\mathcal{R}$.  Moreover, a closed-form value for a lower bound on the mean number of clusters $K$ to form can be obtained as a function of $m$, the number of sampled intensities generated by a point process, $\mathcal{P}$.

Suppose an order statistic is applied to a collection of observations from a point process $\mathcal{P}$ that defines a random field $\mathbf{P}$, as in $\cite{Guyon}$.  After projection from higher dimensional observations into $2$ dimensions, the novelty of the result from $\cite{Murphy}$ is robust cluster membership of each open edge.  In addition, it is proven in $\cite{Murphy}$ that $\mathcal{R}$ is continuous as a function of $m$ and  $K=K(m)$.  Also, $\cite{Murphy}$ provides an analytical method for the calculation of $\mathcal{R}=\mathcal{R}(m,K)$, i.e. no estimation techniques are required.  Thus, we only need to consider $2D$ data, formally giving us reason to consider an image as a random field of sites in a $2D$ integer lattice.

Site states in the lattice are pixel intensities and associated edge states are determined by equivalence of neighboring pixel intensities.  Then, the connectivity radius $\mathcal{R}$ is an integer value as a count of pixels to the left, right, above or below a center pixel, consequently defining a local receptive field in an image.  A local receptive field gives us a calculable method for partitioning an image into overlapping sub-regions when opening and closing edges.  Note that if $\mathcal{R}$ is odd, then the center "pixel" of a receptive field is virtual and sits at the intersection of an even number of "pixels".

\section{Random Fields}
\label{rf}

\subsection{Core Results}
\label{coresec}

As in $\cite{Murphy2}$, imagine a closed, bounded region $\mathcal{B}$ in the $2D$ plane, partitioned by uniformly spaced vertical and horizontal lines, $x = c \in \{1,2,...,C\}$ and $y = r \in \{1,2,...,R\}$, respectively, to form an $R \times C$ array of intersections. At the intersections of the orthogonal lines, a process $\mathcal{P}$ independently generates integer values $i_{(r,c)} \in \mathbb{N}$ at $(r,c) \in \mathbb{Z}^{2}$ according to some probability distribution, $\mathbf{P}$.  We flatten the $2D$ array into a $1D$ vector to enumerate the sites in $\mathcal{B}$ as $1,2,...,d = R \times C$ and say that sites $t,t^{\prime} \in \mathcal{B}$ are $\mathbf{neighbors}$ if the \textit{Euclidean distance} $\|t-t^{\prime}\|_{2}=1$.  Define the sample space of tuples $\Omega = \{(i_{t} \in \mathbb{N})_{t \in \mathcal{B}}\}$ and let $\mathcal{A}$ be a $\mathbf{\sigma-algebra}$ of subsets of $\Omega$ such that $\emptyset,\Omega \in \mathcal{A}$ with $\mathcal{A}$ being closed under finite intersections and countable unions of its elements.

\begin{definition}
\label{def1}
For a fixed site $t \in \mathcal{B}$, define $\mathcal{I}_{t} = \{i_{t} \in \mathbb{N}\}$ to be the set of all its possible state values.  Suppose $\mathcal{X} = \{X_{t}\}_{t=1}^{d}$ is a sequence of random variables such that $X_{t}$ is zero-mean and square integrable on the probability space $(\Omega,\mathcal{A},\mathbf{P})$ and taking values in $\mathcal{I}_{t}$ for all $t = 1,2,...,d$.  For chosen $\epsilon > 0$, an $\mathbf{edge}$ $e_{t \leftrightarrow t^{\prime}} \in \mathbf{E}$ is $\mathbf{open}$ if $\mathbf{P}(|X_{t}-X_{t^{\prime}}| \ge \epsilon) = 0$ whenever $\|t-t^{\prime}\|_{2}=1$.  Otherwise, $e_{t \leftrightarrow t^{\prime}} \in \mathbf{E}$ is $\mathbf{closed}$.
\end{definition}

Generally, an edge $e_{t \leftrightarrow t^{\prime}}$ in the edge space $\mathbf{E}$ that corresponds to the set of neighboring sites $t,t^{\prime} \in \mathcal{B}$ is open only if $|i_{t}-i_{t^{\prime}}|=0$ for $i_{t} \in \mathcal{I}_{t}$ and $i_{t^{\prime}} \in \mathcal{I}_{t^{\prime}}$, requiring $\mathbf{P}(|X_{t}-X_{t^{\prime}}| \ge \epsilon) = 0$ for all $\epsilon > 0$.  However, the choice of $\epsilon > 0$ in def. $(\ref{def1})$ can be a fixed integer value, allowing for some margin of error in a heuristic determination of open edges across $\mathcal{B}$.  Indeed, in $\cite{Geman}$, at each stopping time in the schedule of temperature changes, sites are chosen at random, and its corresponding state value is updated with the opposite value in $\mathbf{S}$.  Then, the margin of error $\epsilon > 0$ allows for a heuristic determination of which changes maximize a posterior distribution according to some chosen criteria.

Suppose $X_{t} = i_{t}$ is a changed state value resulting in an increase in entropy (energy), as measured by a calculation of the probability of configuration of edge states across $\mathcal{B}$.  Then, $\mathbf{P}(|i_{t}-X_{t^{\prime}}| \ge \epsilon) = \mathbf{P}(|X_{t}-X_{t^{\prime}}| \ge \epsilon) = 0$ for $\epsilon > 0$ if and only if $\mathbf{P}(|i_{t}-X_{t^{\prime}}| < \epsilon) = 1$ for $\epsilon > 0$ whenever $\|t-t^{\prime}\|_{2}=1$.  Now, since $\epsilon$ is assumed to be a positive integer, then choosing $\epsilon = 1$ shows that $X_{t^{\prime}} = i_{t}$, except possibly on a set of $\mathbf{P}$-measure zero.  Thus, $\mathbf{P}(|i_{t}-X_{t^{\prime}}| < \epsilon) = 1$ implies the edge $e_{t \leftrightarrow t^{\prime}} \in \mathbf{E}$ is open after the state change and the change is kept with probability $p$ because of the increase in entropy.  With this site replacement heuristic as in $\cite{Geman}$, there is an added benefit of preventing the algorithm from terminating at sub-optimal, local maxima during the search for the maximizing posterior probability distribution from which the sample data are drawn.  In $\cite{Geman}$, it is shown that under the condition $\mathbf{P}(|X_{t}-X_{t^{\prime}}| \ge \epsilon) = 0$, the annealing algorithm converges to the optimal distribution, $\mathbf{P}$.  From $\cite{Grimmett,Grimmett2}$, it is equivalent to the statement that all sites are almost surely connected in a single open cluster of edges at convergence.  Then, at equilibrium, local image intensities have the same values, except on a set of $\mathbf{P}$-measure zero, where symmetries are broken, resulting in isolated clusters of grayscale to reveal objects and features within an image.

\begin{definition}
\label{postdef}
Let $\mathcal{B}_{t} = \{t^{\prime} \in \mathcal{B} : \|t-t^{\prime}\|_{2}=1\}$.  From $\cite{Guyon}$ and $\cite{Geman}$, the Kolmogorov Theorem allows a $\mathbf{random\ field}$ to be defined as the maximal posterior probability distribution $\mathbf{P}$ obtained by applying an annealing and edge state replacement heuristic to the set of single-edge conditionals, $\mathcal{C} = \{\mathbf{P}_{t \leftrightarrow t^{\prime}}:t^{\prime} \in \mathcal{B}_{t} \backslash \{t\}\}$ over $\mathbf{E}$.
\end{definition}

For $t^{\prime} \in \mathcal{B}_{t}  \backslash \{t\}$, we have $t \in \mathcal{B}_{t^{\prime}}  \backslash \{t^{\prime}\}$ so that from $\cite{Guyon}$, $\mathcal{C}$ is a \textbf{specification} of single-edge conditionals of some $\mathbf{P}$.  Therefore, by the Kolmogorov Theorem, a unique $\mathbf{P}$ exists whose single-edge conditionals are in $\mathcal{C}$, up to sets of $\mathbf{P}$-measure zero in the $\sigma$-algebra of subsets of $\mathbf{E}$.  Thus, def. $(\ref{postdef})$ is consistent.

\begin{theorem}
\label{postthm}
For given $\epsilon > 0$, let $\delta=\delta(\epsilon) > 0$ be defined.  Then, the limiting posterior distribution $\mathbf{P}^{\delta}$ is not maximal, if $\mathbf{P}(|X_{t}-X_{t^{\prime}}| \ge \epsilon) = \delta$, whenever $\mathcal{C}$ is the set of single-edge conditionals of $\mathbf{P}$.
\end{theorem}

If, with non-zero probability $\delta > 0$, we allow edges $e_{t \rightarrow t^{\prime}} \in \mathbf{E}$ to be open for $t^{\prime} \in \mathcal{B}_{t} \backslash \{t\}$ when $|i_{t}-i_{t^{\prime}}| > 0$, then

\begin{itemize}
\item Specifying an integer $\epsilon > 0$,
\item With $\delta = \delta(\epsilon) > 0$,
\item And an associated test condition during an annealing process
\end{itemize}

\noindent
defines a \textit{heuristic} for the determination of certain open edges across $\mathcal{B}$, since $\mathbf{P}^{\delta}$ is sub-optimal by thm. $(\ref{postthm})$.  Later, we will see that applying an assumption of \textit{stochastic separability} to the distribution $\mathbf{P}$ is the same as finding $\delta > 0$ such that $\mathbf{P}(|X_{t}-X_{t^{\prime}}| \ge \epsilon) = \delta$, resulting in a sub-optimal posterior, $\mathbf{P}^{\delta}$.

\begin{definition}
\label{weakdef}
For $t \in \mathcal{B}$, define $F_{t} = \mathbf{P}X_{t}^{-1}$ to be the cumulative distribution of $\mathbf{P}$ at site $t$.  Given $\epsilon,\delta > 0$ and the condition $\mathbf{P}(|X_{t}-X_{t^{\prime}}| \ge \epsilon) = \delta$, we say that $\mathbf{P}^{\delta}$ converges weakly to $\mathbf{P}$ as $\delta \rightarrow 0$ (written $\mathbf{P}^{\delta} \xrightarrow{W} \mathbf{P}$) if and only if $F_{t}^{\delta} = \mathbf{P}^{\delta}X_{t}^{-1} \rightarrow \mathbf{P}X_{t}^{-1} = F_{t}$ on $\mathcal{I}_{t}$ as $\delta \rightarrow 0$.
\end{definition}

\begin{theorem}
\label{weakthm}
For chosen $\epsilon > 0$ and $\delta = \delta(\epsilon) > 0$ such that $\mathbf{P}(|X_{t}-X_{t^{\prime}}| \ge \epsilon) = \delta$ as in thm. $(\ref{postthm})$, we have $\mathbf{P}^{\delta} \xrightarrow{W} \mathbf{P}$ as $\delta \rightarrow 0$.
\end{theorem}

Suppose $m > 0$ is an integer such that $m^{2} < d$.  For overlapping $m \times m$ sub-regions of sites $\mathcal{B}_{t}^{m}$ centered at $t \in \mathcal{B}$, suppose an edge $e_{t \leftrightarrow t^{\prime}} \in \mathbf{E}$ has its state updated during annealing.

\begin{lemma}
\label{symlem}
If an update results in broken symmetry (i.e. $|i_{t}-i_{t^{\prime}}| > 0$ with non-zero probability), then for $\epsilon > 0$, there exists $\delta = \delta(\epsilon) > 0$ such that $\mathbf{P}(|X_{t}-X_{t^{\prime}}| \ge \epsilon) = \delta$, where $|i_{t}-i_{t^{\prime}}| \ge \epsilon > 0$.
\end{lemma}

From $\cite{Geman}$, the point process $\mathcal{P}$ that renders an image is thought to independently generate intensities at sites $t \in \mathcal{B}$ according to some probability distribution $\mathbf{P}$.  The dual of the point process is a \textit{line process} $\mathcal{L}$ that generates open edges with non-zero probability in $(0,1)$ between neighboring sites with non-zero probability of having the same pixel intensities.  Independently of the point and line processes, a \textit{stationary} noise process adds blurring and other degradation effects which are to be smoothed during annealing.  Therefore, edges are open or closed with non-zero probability and all edges have non-zero probability of undergoing an update.  If an update results in an increase in entropy, then the update is kept with some probability, $1-p$ for some $p \in (0,1)$.  Otherwise, updates are always kept when entropy decreases.  This algorithm always converges to the equilibrium distribution, provided the annealing schedule is as prescribed in $\cite{Geman}$.  It follows that

\begin{lemma}
\label{updlem}
After an update to the state of an edge $e_{t \leftrightarrow t^{\prime}} \in \mathbf{E}$ in overlapping sub-regions $\mathcal{B}_{t}^{m},\mathcal{B}_{t^{\prime}}^{m} \subset \mathcal{B}$ such that $t^{\prime} \in \mathcal{B}_{t}^{m} \backslash \{t\}$ and $t \in \mathcal{B}_{t^{\prime}}^{m} \backslash \{t^{\prime}\}$, we have $\mathbf{P}(|i_{t}-i_{t^{\prime}}| > 0) > 0$.
\end{lemma}

By an assumption of stochastic separability, partitioning $\mathcal{B}$ into $m \times m$ overlapping sub-regions results in non-zero probability of broken symmetry during annealing.  By lem. $(\ref{symlem})$, given $\epsilon > 0$, there exists $\delta = \delta(\epsilon) > 0$ such that $\mathbf{P}(|X_{t}-X_{t^{\prime}}| \ge \epsilon) = \delta$ for $t^{\prime} \in \mathcal{B}_{t} \backslash \{t\}$.  The next corollary of lem. $(\ref{symlem})$ follows directly from these statements and thms. $(\ref{postthm},\ref{weakthm})$.

\begin{corollary}
\label{stochcor}
For an arbitrary integer $m > 0$ such that $m^{2} < d$, assume the bounded region of sites $\mathcal{B}$ is partitioned into overlapping sub-regions $\mathcal{B}_{t}^{m}$ of size $m \times m$ and centered at $t \in \mathcal{B}$.  If an annealing process is applied to each sub-region, independently of all others, then for given $\epsilon > 0$, the resulting posterior distribution $\mathbf{P}^{\delta}$ is almost surely not maximal under the condition $\mathbf{P}(|X_{t}-X_{t^{\prime}}| \ge \epsilon) = 0$.  As $m^{2} \rightarrow d$, $\mathbf{P}^{\delta}$ converges weakly to $\mathbf{P}$.
\end{corollary}

Implicit in the result of cor. $(\ref{stochcor})$ is the dependence of $\delta$ on the integer $m > 0$ such that $\delta \rightarrow 0$ as $m^{2} \rightarrow d$, yielding the condition $\mathbf{P}(|X_{t}-X_{t^{\prime}}| \ge \epsilon) = 0$ for $t^{\prime} \in \mathcal{B}_{t}^{m} \backslash \{t\}$.  By thm. $(\ref{weakthm})$, $\mathbf{P}$ is maximal following termination of annealing and from $\cite{Grimmett,Grimmett2}$, we know that $\mathbf{P}$ is unique, with all sites in $\mathcal{B}$ almost surely connected by open edges.  Then, also from $\cite{Geman,Grimmett,Grimmett2,Murphy}$, given fixed $\epsilon > 0$, there exists an integer $m_{c} > 0$ with $m_{c}^{2} \le d$ such that under $\mathbf{P}(|X_{t}-X_{t^{\prime}}| \ge \epsilon) = \delta(\epsilon,m)$, we have $\mathbf{P}$ being maximal for all $m > m_{c}$.

\begin{proposition}
\label{critprop}
It follows that $\delta(\epsilon,m) > 0$ for all $m \le m_{c}$ and $\delta(\epsilon,m) = 0$ for all $m > m_{c}$.
\end{proposition}

Next, we will show how to calculate $m_{c} > 0$ for use in determining an optimal size for the overlapping sub-regions, $\mathcal{B}_{t}^{m_{c}}$, centered at $t \in \mathcal{B}$.

\subsection{Critical Size of the Sub-regions}
\label{critsec}

In $\cite{Murphy}$, data points are assumed to be sampled from a high dimensional continuum and bijectively mapped to a bounded, partitioned $2D$ space.  Spatial statistical properties, such as cluster membership, are preserved and can be determined by density estimation techniques.  One such estimation technique in the continuum is nearest-neighbor measurement, whereby Euclidean distance is calculated between data points, with distances below a certain threshold giving an indication of statistical correlation.  Closer distances between data points in the continuum implies greater correlation.

\begin{definition}
\label{distdef}
In partitioned $2D$ space, \textbf{Hamming distance}, $h(t,t^{\prime})$, between sites, $t,t^{\prime} \in \mathcal{B}$ is defined to be the minimum number of edges in a path joining $t$ to $t^{\prime}$.
\end{definition}

In $\cite{Guyon}$, a \textbf{clique} is defined as the set of sites over which a conditional distribution is defined as an element of a specification.  On $\mathcal{C}$, each clique consists only of $2$ sites, $t,t^{\prime} \in \mathcal{B}$ such that $t^{\prime} \in \mathcal{B}_{t} \backslash \{t\}$, so that $h(t,t^{\prime}) = \|t-t^{\prime}\|_{2} = 1$.  And, an edge $e_{t \rightarrow t^{\prime}} \in \mathbf{E}$ is open (or closed) independently of all others, since the point process $\mathcal{P}$ is assumed to generate an intensity at each site $t \in \mathcal{B}$ independently of all others.  Therefore, $\mathcal{B}_{t}^{m} \subset \mathcal{B}$, for a site $t \in \mathcal{B}$ and an integer $m > 0$, defines a clique whose maximal (conditional) posterior distribution $\mathbf{P}_{t}^{m}$ can be computed using annealing and the Kolmogorov Theorem restricted to the specification, $\mathcal{C}_{t}^{m} = \{\mathbf{P}_{t \rightarrow t^{\prime}} \in \mathcal{C}$\ :\ $t^{\prime} \in \mathcal{B}_{t}^{m} \backslash \{t\}\}$.  Again applying the Kolmogorov Theorem to the specification $\{\mathbf{P}_{t}^{m}\}$, it follows from $\cite{Geman,Grimmett,Grimmett2}$ by uniqueness of the maximal posterior distribution that annealing using either specification $\{\mathbf{P}_{t}^{m}\}$ or $\mathcal{C}$ results in the same posterior $\mathbf{P}$, up to sets of $\mathbf{P}$-measure zero.

Define $\partial\mathcal{B}_{t}^{m} = \{t^{\prime\prime} \in \mathcal{B}\backslash\mathcal{B}_{t}^{m}:\min_{t^{\prime} \in \mathcal{B}_{t}^{m}\backslash\{t\}} h(t^{\prime},t^{\prime\prime})=1\}$ to be the boundary of $\mathcal{B}_{t}^{m}$ in $\mathcal{B}$ so that $m+1 = d(t,\partial\mathcal{B}_{t}^{m}) \doteq \min_{t^{\prime\prime} \in \partial\mathcal{B}_{t}^{m}}{h(t,t^{\prime\prime})}$.  As in $\cite{Murphy}$, suppose $K=K(m)$ is the positive square root of a lower bound on the expected number of disjoint clusters of correlated data points.  Then, $m_{c} > 0$ can be computed for fixed $\rho > 0$, determined by the partitioning structure applied to $\mathcal{B}$, by finding an integer $K=K(m)$ that satisfies

\begin{equation}
\label{meq}
\frac{m^{2}}{m^{2}+2m(K-1)^{2}}=\rho.
\end{equation}

For the square lattice partitioning of $\mathcal{B}$, it is proven in $\cite{Kesten}$ that $\rho = 1/2$.  Since $m > 0$ is known, then from $\cite{Murphy}$, $K=K(m)$ can be calculated from eq. $(\ref{meq})$ as the greatest integer not less than the maximum of $2$ and a solution of eq. $(\ref{meq})$ for $K$, which we will call $K_{c}$.  From these statements and $\cite{Murphy}$,

\begin{theorem}
\label{critmthm}
\begin{equation}
\label{critmeq}
\begin{split}
m_{c} = \floor[\bigg]{\frac{m}{K_{c}}}.
\end{split}
\end{equation}
\end{theorem}

From sec. $(\ref{lm})$, $\mathcal{R}=\mathcal{R}(m,K)$ is continuous as a function of both $m$ and $K$.  In addition, it is shown in $\cite{Murphy}$ that $\mathcal{R}$ is monotone decreasing as a function of increasing $m > 0$ and it is monotone increasing as a function of decreasing $K$.  Now, also from sec. $(\ref{lm})$, we know that the critical radius of the uniformly sized sub-regions is given by
\begin{equation}
\label{Req}
\begin{split}
\mathcal{R}_{c} &= \mathcal{R}(m_{c},K_{c}) \\&= d(t,\partial\mathcal{B}_{t}^{m_{c}})-1 \\&= m_{c}
\end{split}
\end{equation}
\noindent
for all $t \in \mathcal{B}$.  From sec. $(\ref{coresec})$, if $r = m_{c} = c$, then by thm. $(\ref{postthm})$, there is non-zero probability of broken symmetry for integer intensities between any two neighboring sites.  This will be important to the imaging problem, as $m > 0$ increases and details a more disconnected state at lower values of $m$ near $m_{c}$, to a more connected state at higher values of $m > m_{c}$, which directly coincides with the result of cor. $(\ref{stochcor})$.  One other important fact to note is that

\begin{theorem}
\label{Rdecaythm}
$\mathcal{R}$ decays exponentially for $K > K_{c}$.
\end{theorem}

Indeed, by appealing to arguments in $\cite{Grimmett,Grimmett2,Murphy}$ again, a bounded sub-region consists of a single open edge so that $K_{c}^{2}$ edges exist within an $R \times C$ image, where consecutive edges almost surely have non-repeating states, as the original bounded sub-regions are all disjoint by construction in $\cite{Murphy}$.  Thus, if we consider an $R \times C$ image as a bounded region within a larger (possibly infinite) image, then without loss of generality, we may assume that all edges are open in the complement of the smaller image within the larger (possibly infinite) image.  As $K > K_{c}$ by assumption, then the left side of eq. $(\ref{meq})$ is less than $1/2$, the critical probability of open edges on a square lattice.

Intuition about "interesting" images leads to a correct conclusion, in this case.  As the average number of detectable, segregated objects increases within the bounds of an image on the whole, or within a smaller portion of an image, thm. $(\ref{Rdecaythm})$ tells us that the likelihood is slim for all of those objects to be related.  This is an important fact to note for object segmentation and classification tasks.

\section{Deep Belief Network (DBN)}
\label{dbnsec}

\subsection{Target Distribution: Local, Intermediate and Long Range}
\label{targsec}

Recall from sec. $(\ref{intro})$ that an image is a point process generating integer intensities in some $R \times C$ integer lattice.  We have seen from cor. $(\ref{stochcor})$ that if local receptive fields are defined in the lattice for $m > 0$ such that $m^{2} < d$, then the resulting posterior distribution will almost surely be sub-optimal following annealing.  Given $\epsilon > 0$, there exists $\delta = \delta(\epsilon,m) > 0$ such that $\mathbf{P}(|X_{t}-X_{t^{\prime}}| \ge \epsilon) = \delta(\epsilon,m)$, requiring non-zero probability of broken symmetry at the end of annealing.  We called this (sub-optimal) equilibrium distribution, $\mathbf{P}^{\delta}$ and showed in thm. $(\ref{weakthm})$ that $\mathbf{P}^{\delta}$ converges weakly to the optimal $\mathbf{P}$ as $\delta \rightarrow 0$.  By prop. $(\ref{critprop})$, $\delta \rightarrow 0$ as $m^{2} \rightarrow d$.

Let $m = m_{c}$ in the set of bounded sub-regions, $\mathbf{B}^{m} = \{\mathcal{B}_{t}^{m}:t \in \mathcal{B}\}$.  We know from sec. $(\ref{coresec})$ that annealing applied to the specification $\{\mathbf{P}_{t}^{m}\}$ leads to the same maximal posterior $\mathbf{P}$ as the specification $\mathcal{C}$, up to sets of $\mathbf{P}$-measure zero.  If we restrict the annealing process to non-overlapping, mutually-disjoint sub-regions, $\mathcal{B}_{t}^{m},\mathcal{B}_{t^{\prime}}^{m} \in \mathbf{B}^{m}$, then

\begin{theorem}
\label{altthm}
For $m \le m_{c}$, the states are almost surely different for every $2$ contiguous edges $e_{t \leftrightarrow t^{\prime}},e_{t^{\prime} \leftrightarrow t^{\prime\prime}} \in \mathbf{E}$ such that $t^{\prime} \in \mathcal{B}_{t}^{m_{c}} \backslash \{t\}$ and $t^{\prime\prime} \in \mathcal{B}_{t^{\prime}}^{m_{c}} \backslash \{t^{\prime}\}$.
\end{theorem}

For $m = m_{c}$, arguments in $\cite{Geman,Grimmett,Grimmett2,Murphy}$ require the collection of edges across any single, disjoint sub-region in $\mathbf{B}^{m_{c}}$ to collectively be open or closed, almost surely, following annealing applied to the specification $\{\mathbf{P}_{t}^{m}\}$.  Then, as in $\cite{Grimmett,Grimmett2}$, without loss of generality, the associated edge space across each disjoint bounded sub-region contains a single edge that is either open or closed, respectively.  From $\cite{Murphy}$ and by thm. $(\ref{critmthm})$, the states are different for the contiguous edges $e_{t \leftrightarrow t^{\prime}},e_{t^{\prime} \leftrightarrow t^{\prime\prime}} \in \mathbf{E}$ associated to disjoint sub-regions, $\mathcal{B}_{t}^{m_{c}},\mathcal{B}_{t^{\prime\prime}}^{m_{c}} \in \mathbf{B}^{m_{c}}$.  After annealing is applied to $\{\mathbf{P}_{t}^{m}\}$, the edges $e_{t \leftrightarrow t^{\prime}} \in \mathbf{E}|_{\mathcal{B}_{t}^{m_{c}}}$ and $e_{t^{\prime} \leftrightarrow t^{\prime\prime}} \in \mathbf{E}|_{\mathcal{B}_{t^{\prime\prime}}^{m_{c}}}$ almost surely have different states, by arguments in $\cite{Grimmett,Grimmett2,Murphy}$.  Thus, after annealing is applied to $\mathcal{C}$, the Kolmogorov Theorem shows the edges $e_{t \leftrightarrow t^{\prime}},e_{t^{\prime} \leftrightarrow t^{\prime\prime}} \in \mathbf{E}$ alternate between sequences of sites $t,t^{\prime},t^{\prime\prime} \in \mathcal{B}$, almost surely. The case $m < m_{c}$ follows by thm. $(\ref{Rdecaythm})$, since $K > K_{c}$ is guaranteed.  From these arguments, we draw the conclusion of thm. $(\ref{altthm})$.

One might look at the arguments from $\cite{Geman,Grimmett,Grimmett2,Murphy}$ and conclude that the preceding paragraph following the statement of thm. $(\ref{altthm})$ offers contradictory elements, since edges over bounded sub-regions almost surely have the same state after annealing is applied to the specification $\{\mathbf{P}_{t}^{m}\}$, while the result is almost surely alternating states over the same set of edges after annealing is applied to $\mathcal{C}$.  However, the difference between the states over all edges after annealing in both cases amounts to the differences in the distributions up to sets of $\mathbf{P}$-measure zero.  Therefore,

\begin{corollary}
\label{coraltthm}
If $m \le m_{c}$, then $\mathbf{P}$ is uniform.
\end{corollary}

That the (sub-optimal) limiting distribution $\mathbf{P}$ is uniform when $m \le m_{c}$ was always guaranteed from its setup in $\cite{Murphy}$.  In the derivation of eq. $(\ref{meq})$, the ideal case of $d > 0$ data points are generated in a $2D$ bounded region to be uniformly and evenly distributed among $K_{c}^{2}$ disjoint clusters, from which thm. $(\ref{critmthm})$ is obtained.  Edge states are the same within each disjoint cluster so that without loss of generality, each cluster contains only one edge, using arguments from $\cite{Grimmett,Grimmett2}$.  Then, no edge state repeats in the original setup, almost surely.

The astute reader will note also that cor. $(\ref{coraltthm})$ corresponds to simulated chemical annealing in a massively uncoupled site state, owing to uniformity at high temperature parameters.  Then, edges are almost surely closed, a guarantee on the square lattice when $\rho \le 1/2$, which is the case, if $m \le m_{c}$ in eq. $(\ref{meq})$.

We note that any pixel classifier, obtained with $m \le m_{c}$ during annealing applied to disjoint, bounded sub-regions of size $m_{c} \times m_{c}$, will give predictions that are no better than a random guess from the uniform distribution.  This eventuality will be shown experimentally in the next section.  An application of the Central Limit Theorem (for fields) $\cite{Grimmett,Guyon,Hogg}$ can be made to show

\begin{corollary}
\label{coraltthm2}
If $m > m_{c}$, annealing ends in a Gaussian field, $\mathbf{P}$.
\end{corollary}

\subsection{Model of the Distribution}

\subsubsection{Theoretical}

We saw in secs. $(\ref{critsec},\ref{targsec})$ that the posterior $\mathbf{P}$, obtained after annealing is applied to the specification $\{\mathbf{P}_{t}^{m}\}$, is the same as the posterior obtained after annealing is applied to $\mathcal{C}$, up to sets of $\mathbf{P}$-measure zero.  If $m \le m_{c}$, then cor. $(\ref{coraltthm})$ requires that $\mathbf{P}$ is uniform, while cor. $(\ref{coraltthm2})$ requires $\mathbf{P}$ to be Gaussian, if $m > m_{c}$.  Yet, there are certain advantages to sizing the bounded sub-regions $\mathbf{B}^{m} = \{\mathcal{B}_{t}^{m}:t \in \mathcal{B}\}$ with $m = m_{c}$ and using the specification $\{\mathbf{P}_{t}^{m}\}$ to obtain the posterior, $\mathbf{P}$.

Since the posterior $\mathbf{P}$ is uniform following annealing when $m = m_{c}$, then its specification (which matches the conditionals of $\mathbf{P}$ up to sets of measure zero) almost surely contains only one element, say $\mathbf{P}_{t}^{m_{c}}$, defined over its bounded sub-region, $\mathcal{B}_{t}^{m_{c}}$, for some fixed $t \in \mathcal{B}$.  Applying the Kolmogorov Theorem to the specification $\mathcal{C}_{t}^{m_{c}}$, we are able to find a joint distribution over $\mathcal{B}_{t}^{m_{c}}$ with its conditionals in $\mathcal{C}_{t}^{m_{c}}$, up to sets of $\mathbf{P}_{t}^{m_{c}}$-measure zero.  Without loss of generality, we will again refer to this joint distribution as $\mathbf{P}_{t}^{m_{c}}$.  We use the Kolmogorov Theorem to extend $\mathbf{P}_{t}^{m_{c}}$ to $\partial\mathcal{B}_{t}^{m_{c}}$.  Then, by thm. $(\ref{weakthm})$ and prop. $(\ref{critprop})$, we obtain $\mathbf{P}$ in cor. $(\ref{coraltthm2})$ by uniqueness guaranteed in $\cite{Grimmett,Grimmett2}$, since $m = m_{c}+1 > m_{c}$ after the extension.

\begin{theorem}
\label{distthm}
For any fixed $t \in \mathcal{B}$, the maximal posterior distribution $\mathbf{P}$ (up to sets of $\mathbf{P}$-measure zero) is the extension of $\mathbf{P}_{t}^{m_{c}}$ to $\partial\mathcal{B}_{t}^{m_{c}}$.
\end{theorem}

We note that the distribution $\mathbf{P}$ defined in thm. $(\ref{distthm})$ provides for the overlapping local receptive fields sought in sec. $(\ref{metsec})$.  Indeed, if we collect the centers $\mathbf{C}^{m_{c}} = \{t \in \mathcal{B} : \mathcal{B}_{t}^{m_{c}} \in \mathbf{B}^{m_{c}}\}$ of each element of the collection of bounded sub-regions, then $\{\mathbf{P}_{t}^{m_{c}} : t \in \mathbf{C}^{m_{c}}\}$ is an overlapping specification that produces $\mathbf{P}$ (up to sets of $\mathbf{P}$-measure zero) after annealing on $\mathcal{C}_{t}^{m_{c}}$ is extended to $\partial\mathcal{B}_{t}^{m_{c}}$ for each $t \in \mathbf{C}^{m_{c}}$.  By thm. $(\ref{distthm})$, this overlapping specification almost surely contains exactly one element, namely $\mathbf{P}$ itself, so that $\mathbf{P}$ must be our distribution over the whole of $\mathcal{B}$.  By thm. $(\ref{altthm})$, every site in $\mathcal{B}_{t}^{m}$ is almost surely connected in a contiguous sequence of open edges when $m > m_{c}$.  Then, an easy consequence of thm. $(\ref{distthm})$ is

\begin{corollary}
\label{distcor}
Fix $\epsilon > 0$ and suppose $m > m_{c}$.  Then, every site $t \in \mathcal{B}$ is almost surely connected by contiguous open edges to $m$ sites $t^{\prime} \in \mathcal{B} \backslash \{t\}$ under the condition $\mathbf{P}(|X_{t}-X_{t^{\prime}}| \ge \epsilon) = 0$.
\end{corollary}

Note that at least one open edge connects the interior of $\mathcal{B}_{t}^{m_{c}}$ to $\partial\mathcal{B}_{t}^{m_{c}}$ by thm. $(\ref{distthm})$, since $m > m_{c}$, so that $\partial\mathcal{B}_{t}^{m_{c}}$ captures long-range dynamics of $\mathbf{P}$ by cor. $(\ref{distcor})$, which comports with theory in $\cite{Grimmett,Grimmett2}$.  Since the order of $\{\mathbf{P}_{t}^{m_{c}} : t \in \mathbf{C}^{m_{c}}\}$ is almost surely $1$ by thm. $(\ref{distthm})$, prop. $(\ref{critprop})$ and cor. $(\ref{coraltthm2})$, then

\begin{theorem}
\label{statthm}
$\mathbf{P}$ is stationary under the condition, $\mathbf{P}(|X_{t}-X_{t^{\prime}}| \ge \epsilon) = 0$, whenever $m > m_{c}$.
\end{theorem}

\subsubsection{Practical}
\label{pracsec}

From the construction in $\cite{Murphy}$, each disjoint sub-region of $m_{c}^{2}$ intensities generated by the point process $\mathcal{P}$ is separated from its neighboring sub-region by a shared boundary of site states that almost surely form a closed edge in one sub-region and an open edge in the other sub-region, by thm. $(\ref{altthm})$.  Yet, the constructed boundary is virtual in practice, when learning the distribution of actual $2D$ images from annealing applied to the specification $\{\mathbf{P}_{t}^{m_{c}} : t \in \mathbf{C}^{m_{c}}\}$, since the boundary of a given sub-region has non-empty intersection with its neighboring sub-regions.  Therefore, within the context of distributions of pseudo images generated by numerically encoding text as integer values from $\{0,1,2,...,255\}$, the sets of $\mathbf{P}$-measure zero in thm. $(\ref{distthm})$ are insignificant and can be ignored, as only one element of the specification is needed to learn the distribution of a $2D$ text array.  However, in the contexts of object detection and image segmentation, these zero-measure sets can not be ignored since actual objects may account for the differences.  This requires each distribution within $\{\mathbf{P}_{t}^{m_{c}} : t \in \mathbf{C}^{m_{c}}\}$ to be learned in this context.

We simulate the same edge being open in one neighboring sub-region and closed in another by decoupling and parallelizing the learning of each distribution in the specification, $\{\mathbf{P}_{t}^{m_{c}} : t \in \mathbf{C}^{m_{c}}\}$.  Then, each annealing is independently applied to the extension of $\mathbf{P}_{t}^{m_{c}}$ to $\partial\mathcal{B}_{t}^{m_{c}}$ for each $t \in \mathbf{C}^{m_{c}}$.

For image reconstruction, the interior $\mathcal{B}_{t}^{m_{c}}$ of each (extended) sub-region is replaced by a maximum likelihood estimate calculated from boundary elements.  This comports with theory from $\cite{Guyon}$, as the dynamics of the equilibrium distribution $\mathbf{P}$ are dominated by the boundary $\partial\mathcal{B}_{t}^{m_{c}}$ of each extended sub-region, since $m > m_{c}$ after extension.  The Kolmogorov Theorem requires the resulting posterior distribution to be the same as in thm. $(\ref{distthm})$, up to sets of $\mathbf{P}$-measure zero.

That $\mathbf{P}$ is stationary is guaranteed by thm. $(\ref{statthm})$.  However, from the construction, the conditionals of $\mathbf{P}$ are in $\{\mathbf{P}_{t}^{m_{c}} : t \in \mathbf{C}^{m_{c}}\}$, up to sets of $\mathbf{P}$-measure zero, after annealing.  Making another appeal to arguments in $\cite{Grimmett,Grimmett2}$, we may consider each extended bounded sub-region to consist of a single open edge after annealing.  Yet, the extended bounded sub-regions overlap, so that the single edges are also open within the interior of each of its neighboring sub-regions.  Therefore, by construction, all sub-regions are almost surely connected by a contiguous sequence of open edges after annealing, which requires $\mathbf{P}$ to be stationary.

\subsubsection{Implementation}
\label{impsec}

From $\cite{Murphy2}$, a DBN is a layered computational exploitation of the connected graph structure in $\mathbf{E}$ over $\mathbf{B}^{m}$, associated to an overlapping conditional specification $\{\mathbf{P}_{t}^{m} : t \in \mathbf{C}^{m}\}$, for some $m > 0$.  Let $\mathbf{w} \in \mathbf{W}$ be a configuration of edge states over $\mathbf{E}$.  With $\mathbf{w}$ as inputs, hidden layers between the input layer and the final output layer successively estimate parameters to minimize the energy function, $H(\mathbf{x},\beta)$.  Each hidden layer is itself a $3$-layered computational abstraction of the output from the previous layers, consisting of an input, a hidden and an output layer to further refine the estimated parameters.  In each hidden layer in the $3$-layered networks, a series of computational nodes offer multiple local approximations that detail a belief about the global nature of the equilibrium distribution obtained by estimating parameters over local configurations $\mathbf{w}|_{\mathcal{B}_{t}^{m}}$ for $\mathcal{B}_{t}^{m} \in \mathbf{B}^{m}$ and $t \in \mathbf{C}^{m}$, with fixed configurations, $\mathbf{w}|_{\mathcal{B}\backslash\mathcal{B}_{t}^{m}}$.

For the main network, the first hidden layer takes all the data as input, while in the second level there are sibling branches designed to learn the distribution from which the intensity process $\mathcal{P}$ samples, by extracting all information from each output of nodes in the previous layer.  Likewise, the next hidden layer consists of binary branches to learn the line process, $\mathcal{L}$, using all information from the previous layer.  In this way, each layer in the network is fully connected when learning the distribution of intensities and its dual.  Successive layers alternate in this fashion.

Let $J$ represent the layer of the DBN.  For each $3$-layered hidden network (heretofore termed a Restricted Boltzmann Machine (RBM)) containing sibling branches given by each of the $m$ rows of $m$ intensity inputs over $\mathcal{B}_{t}^{m} \in \mathbf{B}^{m}$ and $t \in \mathbf{C}^{m}$, its corresponding RBM consists of $2^{(J-2)/2} \times m^{J/2}$ computational nodes in a hidden layer.  The Gibbs distribution is its activation function, with an identity function at its output layer. For each layer consisting of binary child branches that follow sibling layers, its corresponding RBM follows a similar setup, yet with $2^{(J-1)/2} \times m^{(J-1)/2}$ computational nodes in a hidden layer.  We compile the DBN with a categorical cross entropy loss and adam optimizer.\\

\begin{algorithm}[H]
\SetAlgoLined
 Add a dense layer with tensor shape accepting $m$ columns and any number of rows … a rectified linear unit activation function … with $m$ outputs\;
 Initialize $odim = 2$\;
 \For{$J$ in the range of values from $1$ to $m$}{
  Define input $dim = 2 \times odim$\;
  Add a dense layer with tensor shape accepting input $odim$ columns and any number of rows … a scaled exponential linear unit (Gibbs distribution) activation function … with $dim$ outputs\;
	Define output $odim = 2 \times dim$\;
	Add a dense layer with tensor shape accepting input $dim$ columns and any number of rows … a scaled exponential linear unit (Gibbs distribution) activation function … with $odim$ outputs\;
 }
 Add a dense layer with tensor shape accepting $odim$ columns and any number of rows … scaled exponential linear unit (Gibbs distribution) activation function … with $K_{c}$ outputs\;
 \caption{DBN}
\end{algorithm}

\subsubsection{Example: Structure}
Suppose $m = 2$.  We note that the $2 \times 2$ conditional specification only contains one element by thm. $(\ref{distthm})$.  To learn the uniform distribution guaranteed by cor. $(\ref{coraltthm})$, the structure of the DBN from $\cite{Murphy2}$ takes the form

\begin{figure}[H]
\centering
\includegraphics[scale=0.75]{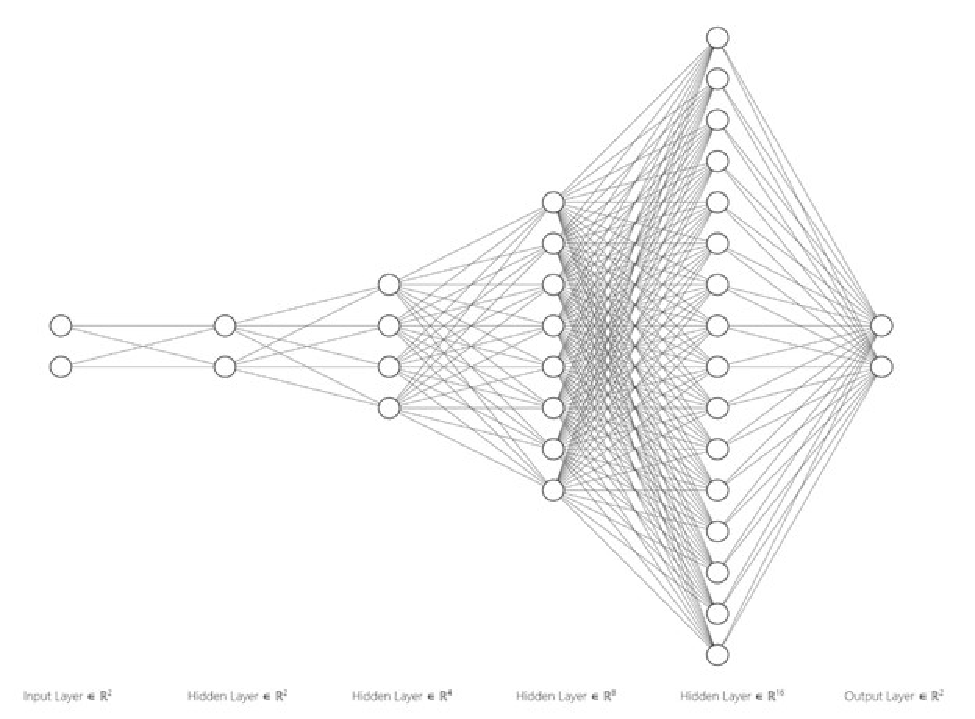}
\caption{DBN with $m = 2$ from $\cite{Murphy2}$}
\label{dbnfig}
\end{figure}

From section "Approach" in $\cite{Murphy2}$, we will have $2$ sibling layers and $2$ binary layers that flow from/to the sibling layers, in an alternating fashion such that the structure of the network consists of $2$ distinct RBMs comprised of layers $(1-3)$ and layers $(3-5)$. The final output layer takes inputs from the output of the final RBM so that the structure of the DBN consists of $2$ RBMs followed by an output layer. From the same section, we know that the number of nodes in each of the sibling layers is $2^{0} \times 2$ and $2 \times 2^{2}$. Likewise, in the binary layers, the number of nodes is $2 \times 2$ and $2^{2} \times 2^{2}$.

Note that layer $(1)$ is the main input to the network and also the input layer of the first RBM.  Layer $(3)$ is the output layer of the first RBM and also the input layer of the second RBM.

\subsubsection{Example: Embedding}
\label{embsec}

Let $m = 2$ and fix $t \in \mathbf{C}^{m}$.  Then, fig. $(\ref{dbnfig})$ details a network structure where $K_{c} = 2$ at the output layer.  In eq. $(\ref{meq})$, we solve for $K = 2$ and label this value $K_{c}$.  Also, from eq. $(\ref{critmeq})$, we have $m_{c} = 1$ so that the size of the bounded region $\mathcal{B}_{t}^{m_{c}}$ is $2 \times 2$, with the center "pixel" of $\mathcal{B}_{t}^{m_{c}}$ being virtual.  By cor. $(\ref{coraltthm})$ we learn the uniform distribution of the image with annealing only being applied to $\mathbf{P}_{t}^{m_{c}}$ over $\mathcal{B}_{t}^{m_{c}}$.  By cor. $(\ref{coraltthm2})$, we learn the equilibrium distribution of the image by applying annealing to a sliding $3 \times 3$ kernel defining the specification, $\{\mathbf{P}_{t}^{3} : t \in \mathbf{C}^{3}\}$ over $\mathbf{B}^{3}$, where the kernel slides $2$ pixels to the left, right, upward or downward to provide overlap.  The reader should note that the $3 \times 3$ kernel is common in image feature extractions using convolutional layers.  Cor. $(\ref{coraltthm2})$ gives theoretical basis for this choice in binary classification and regression problems by an embedding of higher dimensional data into $2D$ space before the $2 \times 2$ input layer, and by noting that regression partitions output into $K_{c} = 2$ classes by default.

\subsubsection{Example: Compression}
\label{compsec}

From $\cite{Murphy2}$, a DBN with the structure given in fig. $(\ref{dbnfig})$ efficiently outputs an equilibrium distribution from uniformly distributed inputs.  By the Kolmogorov Theorem, the equilibrium distribution $\mathbf{P}$ obtained by annealing applied to the $3 \times 3$ specification gives rise to $\mathbf{P}|_{\mathcal{B}_{t}^{2}}$ from the $2 \times 2$ specification, up to sets of $\mathbf{P}|_{\mathcal{B}_{t}^{2}}$-measure zero in the edge space, $\mathbf{E}|_{\mathcal{B}_{t}^{2}}$.

Fix $t \in \mathbf{C}^{2}$.  By thm. $(\ref{distthm})$, $\mathbf{P}$ is the extension of $\mathbf{P}_{t}^{2} = \mathbf{P}|_{\mathcal{B}_{t}^{2}}$ to $\partial\mathcal{B}_{t}^{2}$ and by cor. $(\ref{coraltthm})$, $\mathbf{P}_{t}^{2}$ is uniform.  Therefore, by thm. $(\ref{distthm})$, we generate $\mathbf{P}$ from image intensities by uniformly sampling $2 \times 2 = 4$ values in $\{0,1,2,...,255\}$ to represent the (uniform) conditional distribution $\mathbf{P}_{t}^{2}$.  Note that for each $t \in \mathbf{C}^{2}$, the boundary $\partial\mathcal{B}_{t}^{2}$ in the $3 \times 3$ specification contains $5$ pixels.  To learn the equilibrium distribution $\mathbf{P}$ from the $3 \times 3$ specification, we successively append $5$ pixels to the uniform sample as input to the DBN, for each $t \in \mathbf{C}^{3}$.

Overlap between bounded sub-regions is generated by the common uniform sample.  Only the uniform sample, plus $5$ pixels from each disjoint region in $\mathbf{C}^{3}$, is required to encode the distribution of intensities and edge states in the image, leading to a method for compression and reconstruction with $\mathbf{P}$.

\section{Results}

\subsection{Problem Statement}

Unlike supervised methods for object detection and segmentation $\cite{Girshick,Girshick2,Gkioxari,He,Lin,Liu,Redmon,Ren}$ where images in a training set have been annotated to indicate specific regions containing particular objects, edge detection methods such as the HOG heuristic seek to delineate the boundary (or negative space) between objects.  Yet, from sec. $(\ref{motisec})$, we know that edge detection methods often require normalizing transformations and density estimation techniques to determine statistical correlation of vectors forming edges.  And, it is closed edges (found between normalized regions of statistically correlated vectors in an image) that delineate negative space between different objects.

In the intervening sections, we described an annealing heuristic that can be applied to individual images to find the negative space between objects of interest in an image, along with properties of the probability distribution governing what is observed.  From sec. $(\ref{lm})$, we know that the result of an annealing process applied to pixel intensities in an image is a posterior probability distribution (defined on the edge space) that has certain properties, depending on the size of uniformly-sized sub-regions within an image.  From sec. $(\ref{dbnsec})$, we know how to construct a DBN to model the annealing process and the next sections show how to use the DBN to estimate the posterior distribution, while also showcasing the mathematical results.

\subsection{DBN}

\begin{figure}[H]
\centering
\includegraphics[scale=0.15]{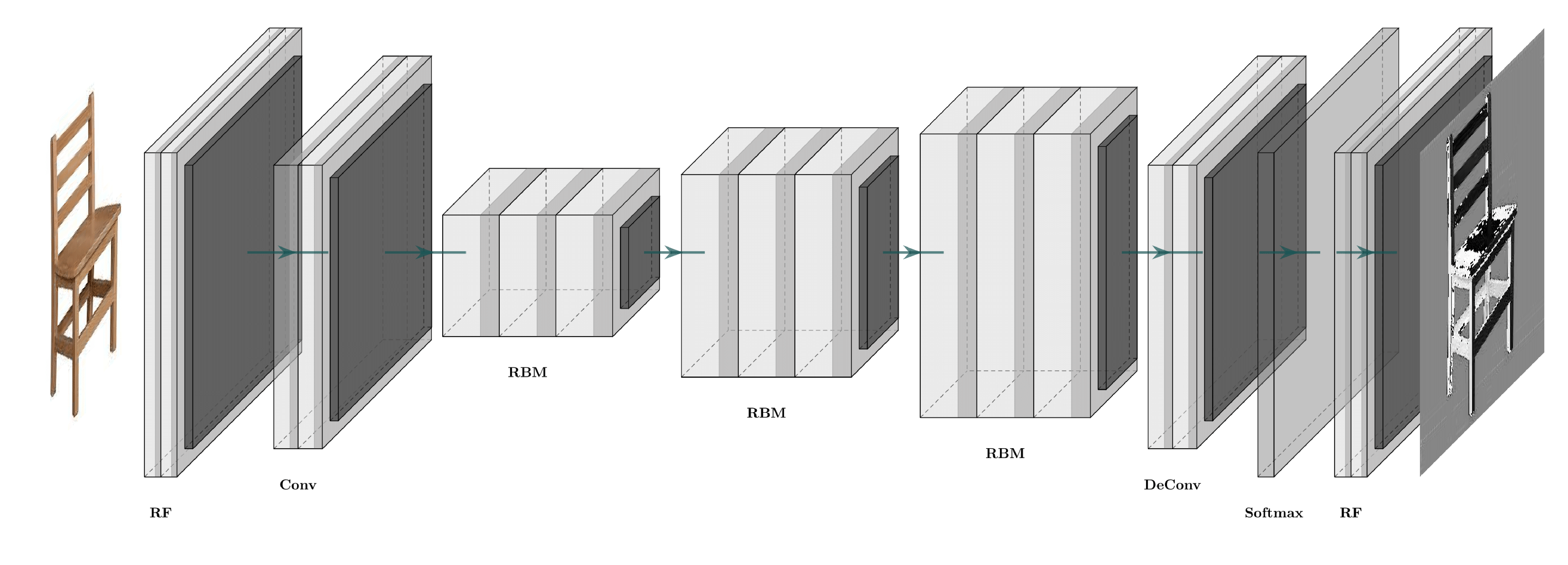}
\caption{Convolutional DBN with an input layer, random field (RF) image disintegration (by stochastic separability), (de) convolution and pooling/upsampling layers, stacked RBMs, a compressed representational layer, RF reintegration and an output layer}
\end{figure}

An image is partitioned into $(m_{c}+1) \times (m_{c}+1)$ sub-regions using stochastic separability from secs. $(\ref{metsec},\ref{rf})$ that is guaranteed by linearity of the convolutional operator and its Gaussian kernel initializer in the practical implementation of the DBN from secs. $(\ref{pracsec},\ref{impsec})$.  Each sub-region proceeds through a series of energy-based, restricted Boltzmann machine layers, which together forms the DBN described in sec. $(\ref{impsec})$.  Deconvolution (convolution transpose) takes place, after which a dense layer supplies prediction probabilities of neighboring pixel intensities in the $(m_{c}+1) \times (m_{c}+1)$ sub-regions for edge creation.  Lastly, the sub-regions (originally partitioned using results from secs. $(\ref{intro},\ref{rf})$) are reintegrated into the original image dimensions and proposals for regions of interest (RP) are determined.

Recall from secs. $(\ref{intro},\ref{rf})$ that an annealing process is designed to open and close edges between neighboring pixels to ensure that a globally optimum posterior distribution is attained.  As such, after equilibrium is attained, regions within an object are smoothed into open edges, resulting in local constancy of intensity, while regions between objects are "hardened" into closed edges, highlighting the greatest changes in intensity.  To accomplish the same feat, we note that within each $(m_{c}+1) \times (m_{c}+1)$ region, the DBN is designed to output probabilities of intensities for filtering, through substitution of the intensity with greatest probability to each of its neighbors.  In this way, edges are open and closed across overlapping $(m_{c}+1) \times (m_{c}+1)$ regions to simulate one facet of the annealing process.  After equilibrium is attained, closed edges are identified by differencing the equilibrium image and the original image.  To account for relative small changes in intensity in localized regions, a threshold is defined, below which, a change is considered to be of zero intensity.  Then, the integrity of an object's interior intensities is maintained, while its boundary is highlighted in the difference image, delineating the RP.

\begin{theorem}
\label{epsilonthm}
For fixed $m > 0$ such that $m^{2} \le d$, the threshold $\epsilon > 0$ in the condition $\mathbf{P}(|X_{t}-X_{t^{\prime}}| \ge \epsilon) = \delta(\epsilon,m)$ has a critical value $\epsilon_{c} > 0$ such that $\mathbf{P}(|X_{t}-X_{t^{\prime}}| \ge \epsilon) = 0$ whenever $\epsilon_{c} \geq \epsilon > 0$.
\end{theorem}

Thm. $(\ref{epsilonthm})$ is just a statement that differences in intensities of a series of connected, open edges in small, bounded regions are almost surely within a defined range not larger than $\epsilon_{c} > 0$.  Indeed, thm. $(\ref{epsilonthm})$ is a consequence of duality between site (intensity) and bond (edge) percolation models on the $2D$ lattice by $\cite[Thm.\ (1.11)]{Grimmett}$, requiring the existence of a critical edge probability, $\delta_{c}=\delta(\epsilon,m_{c})$ for fixed $\epsilon > 0$ (and critical difference intensity, $\epsilon_{c} \geq \epsilon > 0$).  Yet, the utility in the theorem lies in its use for delineating RP by way of smoothing small regions of changing pixel intensities where the differences in intensity are within $\epsilon > 0$.

Suppose annealing is applied to an image and its posterior distribution of intensities is obtained.  Further suppose the smoothed image is obtained by sampling the posterior with intensities at sites shifted in the $2D$ plane by Hamming distance $1$ from their position in the original image.  If $t^{\prime}$ is a permuted counterpart site in the smoothed image associated to a site $t$ in the original image such that $h(t,t^{\prime})=1$, then $\mathbf{P}(|X_{t}-X_{t^{\prime}}| \ge \epsilon) = 0$ whenever $\epsilon_{c} \geq \epsilon > 0$, almost surely.

By thm. $(\ref{epsilonthm})$, RP are delineated by increasing integer intensities $\epsilon \in [0,\epsilon_{c}+1]$ in the difference image since $\delta(\epsilon,m) \rightarrow 0$ as $\epsilon \rightarrow (\epsilon_{c}+1)^{-}$, resulting in transitions from open edges to closed edges at the boundary of an object.  If we set all intensities within $[0,\epsilon_{c}]$ to zero in the difference image, then boundaries of objects can be found by identifying all white-shifted intensities.  Alternatively, we can white-shift the entire image, and identify white-shifted objects, with darker boundaries.

From cor. $(\ref{coraltthm})$, annealing ends in a sub-optimal, uniform distribution when $m \le m_{c}$.  From thm. $(\ref{distthm})$, each upper left $m_{c} \times m_{c}$ sub-region within an $(m_{c}+1) \times (m_{c}+1)$ sub-region of the image is uniform and disjoint from all other sub-regions in the image, after annealing is complete.  RP are then disjoint, requiring merging through boundaries between sub-regions.  It is also an easy consequence of thm. $(\ref{distthm})$ that all RP are almost surely connected, when $m > m_{c}$, at which point, the DBN has attained the equilibrium distribution, which is verified in practical examples.

\subsection{Segmentation and Detection}

\subsubsection{Discussion}

\begin{figure}[H]
\centering
\label{origfig}
  \begin{subfigure}{\linewidth}
  \includegraphics[width=.20\linewidth]{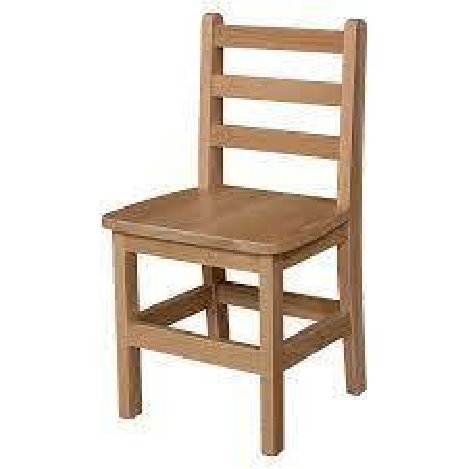}\hfill
  \includegraphics[width=.20\linewidth]{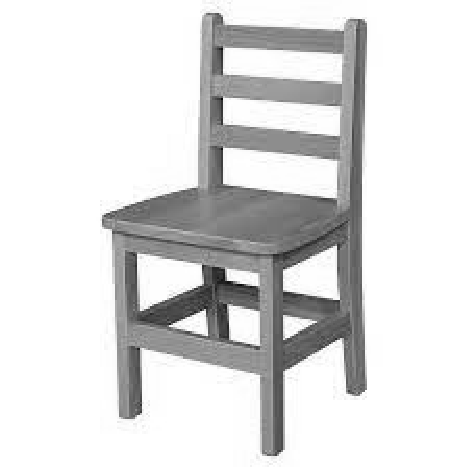}
	\end{subfigure}
  \caption{Original input image before and after conversion to $8$-bit grayscale}
	\label{brown_gray_chairs}
\end{figure}

The original input image is converted to $8$-bit grayscale and its critical radius from thm. $(\ref{critmthm})$, eq. $(\ref{Req})$ and prop. $(\ref{critprop})$ is $m_{c}=2$.  The intensity threshold $\epsilon > 0$ is allowed to monotonically increase over values in the set, $\{70,80,90,...,140\}$, with the object being increasingly segmented from its background by white-shifted pixels as a consequence of thm. $(\ref{epsilonthm})$.  As we will see, the DBN that mimics the annealing process of the modified Metropolis algorithm from sec. $(\ref{metsec})$ also segments an object in an image that is reconstructed from a compressed variant of its original grayscale image.

\subsubsection{Images}

\begin{figure}[H]
\centering
  \begin{subfigure}{\linewidth}
  \centering
  \includegraphics[width=.1\linewidth]{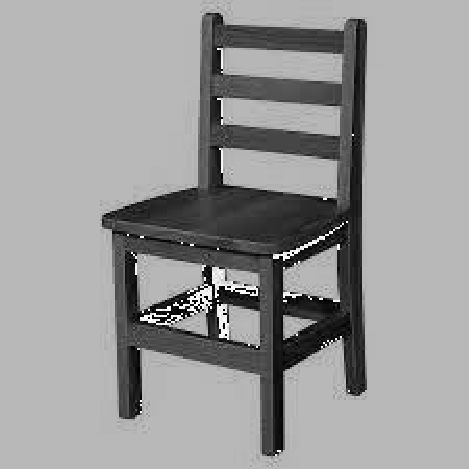}
  \includegraphics[width=.1\linewidth]{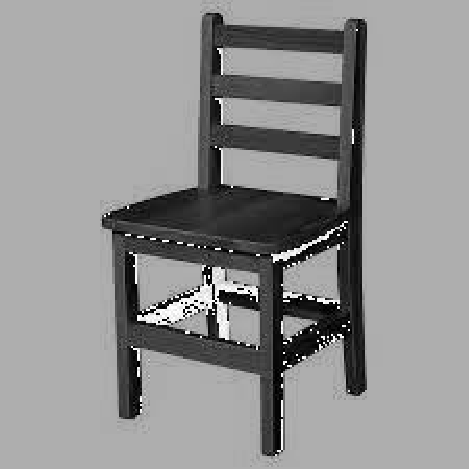}
  \includegraphics[width=.1\linewidth]{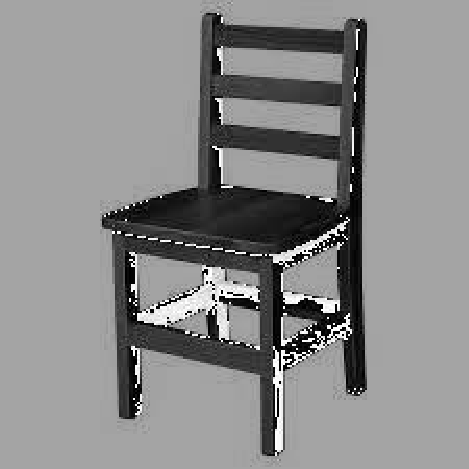}
  \includegraphics[width=.1\linewidth]{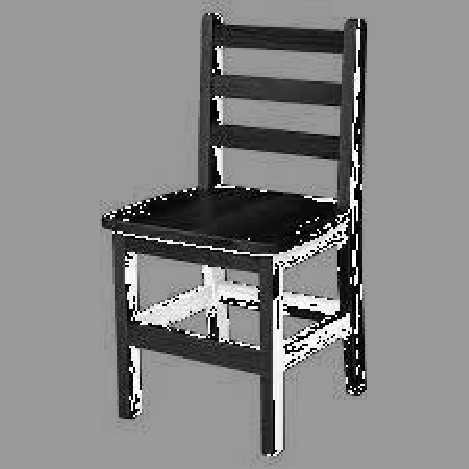}
  \includegraphics[width=.1\linewidth]{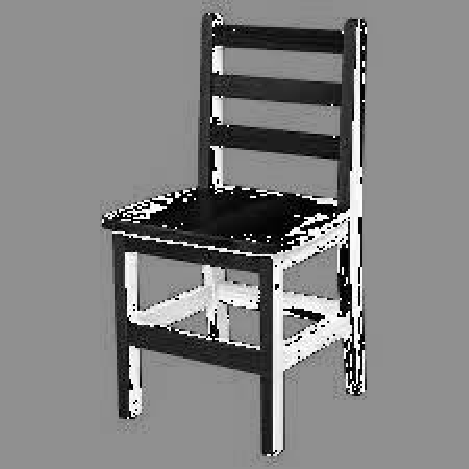}
  \includegraphics[width=.1\linewidth]{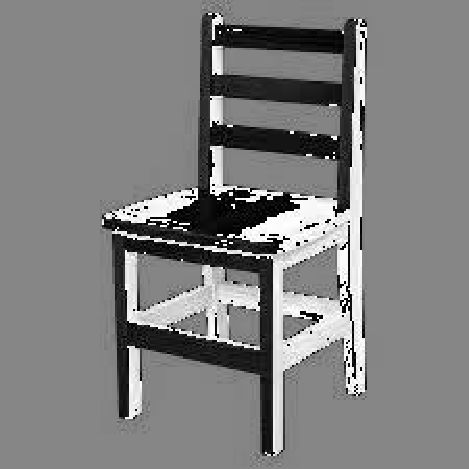}
  \includegraphics[width=.1\linewidth]{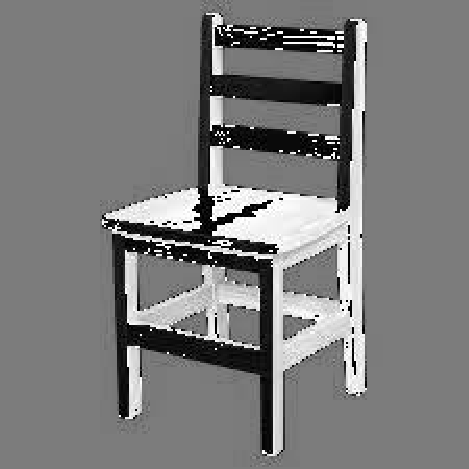}
  \includegraphics[width=.1\linewidth]{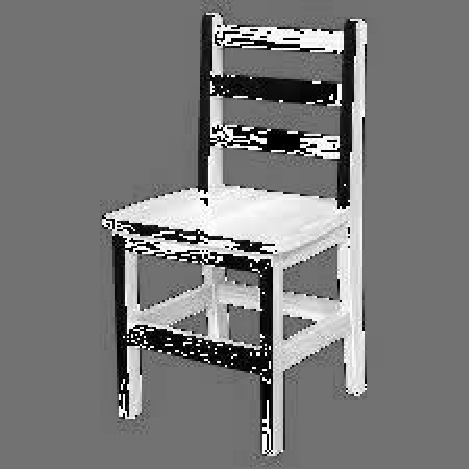}
	\end{subfigure}
  \caption{Object detection with increasing $\epsilon \in \{70,80,90,...,140\}$ in thm. $(\ref{epsilonthm})$}
\label{allchair70_140}
\end{figure}

Consider the first image in the sequence of images in figs. $(\ref{allchair70_140})$, which corresponds to $\epsilon = 70$.  The lower right back leg is starting to be covered in white-shifted pixels to segment the chair from its background.  This serves as experimental evidence that $\epsilon_{c} \le 70$.  As $\epsilon \rightarrow 140$, the chair is increasingly segmented by white-shifted pixels, indicating more open edges are being created in the interior of bounded sub-regions that comprise RP, until eventually the whole chair is segmented.

\subsection{Compression and Reconstruction}

\subsubsection{Discussion}

From cor. $(\ref{coraltthm})$ and thm. $(\ref{distthm})$, each upper left $m_{c} \times m_{c}$ sub-region within an $(m_{c}+1) \times (m_{c}+1)$ sub-region of the image is replaced by the same set of $m_{c}^{2}$ uniformly distributed values.  The substitution values are randomly sampled from $\{0,1,2,...,255\}$ to create a compressed image from the original.  To reconstruct a representation of the original image using the compressed image, the $2D$ Markov property from $\cite{Guyon}$ allows us to reconstruct the upper left $m_{c} \times m_{c}$ sub-region with a maximum likelihood estimate calculated from intensities in its boundary within the $(m_{c}+1) \times (m_{c}+1)$ sub-region.

The compressed image is uniform by cor. $(\ref{coraltthm})$.  Two statistical tests allow for the measurement of normality of intensities in the reconstructed image and how closely its reconstruction reproduces the distribution of intensities of the original image.  Furthermore, if an inference of normality of the reconstructed image can be made, then with the same confidence, we can also infer that the DBN mimics the annealing process of the modified Metropolis algorithm in sec. $(\ref{metsec})$ by thms. $(\ref{distthm},\ref{statthm})$.

The reconstructed image is flattened into a $1D$ vector of length $d = R \times C$.  First, the Shapiro-Wilke (SW) test from $\cite{Shapiro}$ is applied to the resulting vector for comparison to its ordered variant and a test statistic $S$, for some significance level $\alpha \in (0,1)$, measures the difference in energy, $H$.  A value of the test statistic $S > 1 - 2\alpha$ affirms the null hypothesis.  Second, the Kullback-Leibler divergence (KL) test is a measure of the number of relative changes required to intensities of one sample distribution to conform them to another sample distribution.  Values of the KL test statistic fall in the range $[0,\infty)$, with values closest to zero indicating highest correlation between the distributions of intensities.

\subsubsection{Images}

\begin{figure}[H]
\centering
  \begin{subfigure}{\linewidth}
  \includegraphics[width=.1\linewidth]{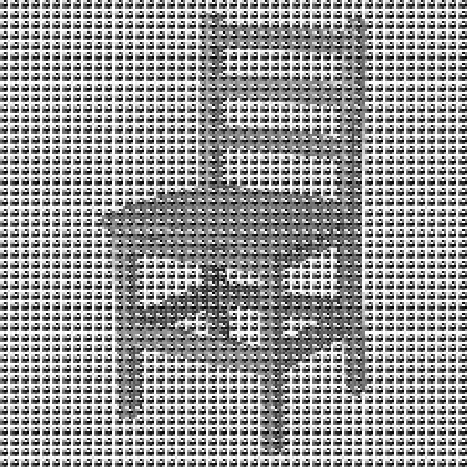}\hfill
  \includegraphics[width=.1\linewidth]{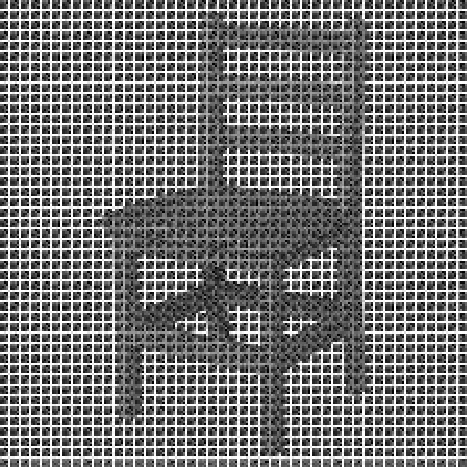}
  \caption{Compression and reconstruction is the same, independent of $\epsilon > 0$ and uniform values substituted in the $m_{c} \times m_{c}$ sub-regions}
	\label{chairs70}
	\end{subfigure}
  \begin{subfigure}{\linewidth}
  \includegraphics[width=.1\linewidth]{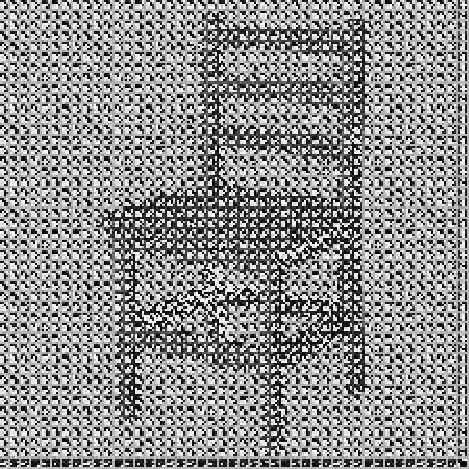}\hfill
  \includegraphics[width=.1\linewidth]{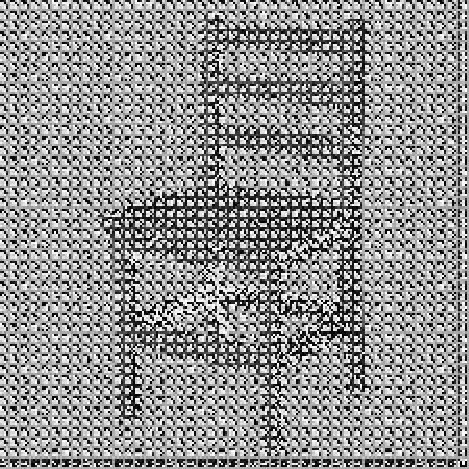}\hfill
  \includegraphics[width=.1\linewidth]{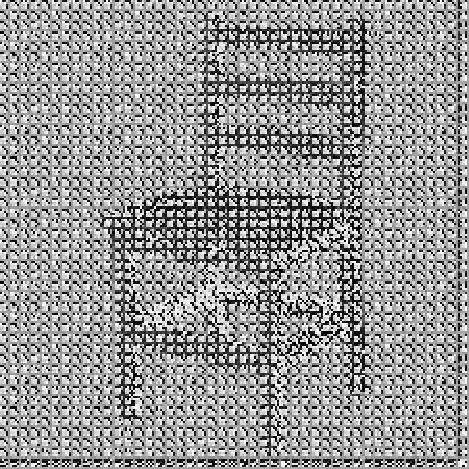}\hfill
  \includegraphics[width=.1\linewidth]{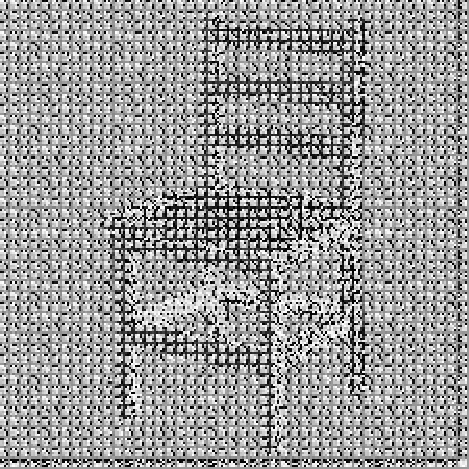}\hfill
  \includegraphics[width=.1\linewidth]{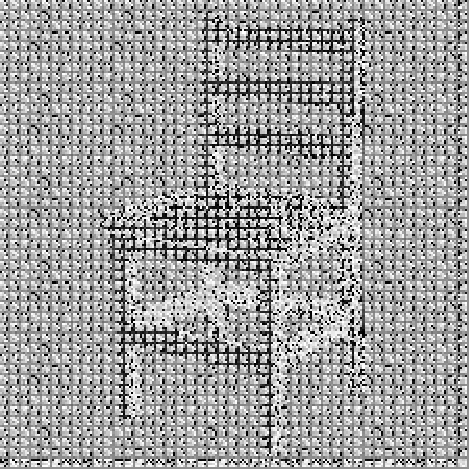}\hfill
  \includegraphics[width=.1\linewidth]{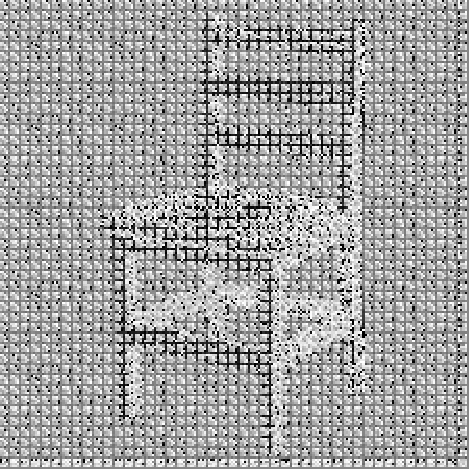}\hfill
  \includegraphics[width=.1\linewidth]{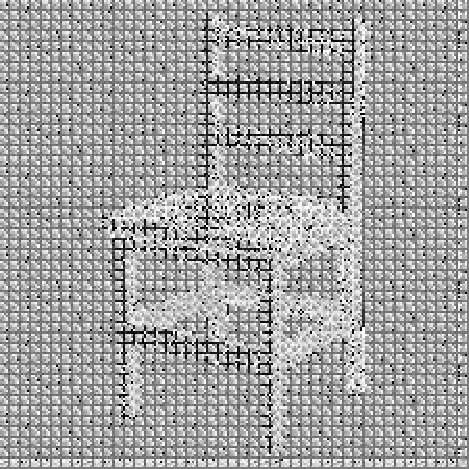}\hfill
  \includegraphics[width=.1\linewidth]{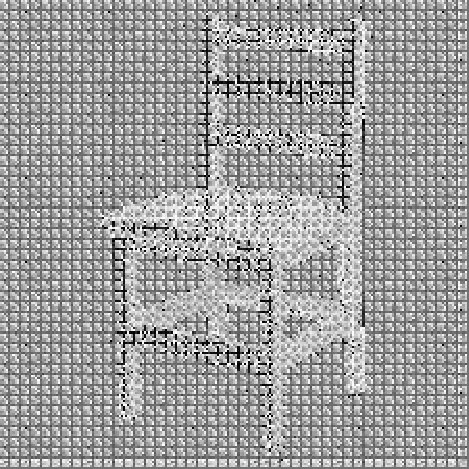}
  \caption{Segmentation using the reconstructed image, with increasing $\epsilon \in \{70,80,90,...,140\}$ in thm. $(\ref{epsilonthm})$}
	\label{allchairs70_140}
	\end{subfigure}
  \caption{Compression, reconstruction and segmentation images, with increasing $\epsilon \in \{70,80,90,...,140\}$ in thm. $(\ref{epsilonthm})$}
	\label{allchairs}
\end{figure}

Figs. $(\ref{chairs70})$ are the compressed and reconstructed variants associated to figs. $(\ref{brown_gray_chairs})$.  Figs. $(\ref{allchairs70_140})$ are obtained after the DBN is applied to the reconstructed image from figs. $(\ref{chairs70})$.  The compressed and reconstructed images are largely unchanged, as they are independent of the intensity threshold $\epsilon > 0$ and independent of the set of uniform values used to substitute each upper left $m_{c} \times m_{c}$ sub-region within each $(m_{c}+1) \times (m_{c}+1)$ sub-region of the original image.

That the compressed and reconstructed images are independent of $\epsilon > 0$ and the uniform substitutions is borne out in the calculation of the KL values for each $\epsilon \in \{70,80,90,...,140\}$, which are constantly $122.7171$.  However, the DBN performs the annealing process to find the equilibrium distribution $\mathbf{P}$ of pixel intensities in the original image and the DBN explicitly depends upon the choice of $\epsilon > 0$.

Note that as $\epsilon > 0$ increases, more white-shifted pixels highlight the object in the reconstructed image.  Indeed, closed edges in the boundary of the object are detected by the DBN, as well as open edges in the interior of the object are detected by the DBN.  Thus, when the DBN is applied to the reconstruction and the original image, the KL values decrease in column $E$ of tab. $(\ref{table:klswtab})$ for increasing $\epsilon > 0$, showing experimental evidence that the equilibrium distribution of the reconstruction intensities becomes a better approximation of the true equilibrium distribution intensities of the original image.

SW statistics in column $S$ of tab. $(\ref{table:klswtab})$ allow an inference of normality after the DBN is applied to the reconstruction, with increasing confidence that tends to $1.0$ as $\epsilon > 0$ increases.  For SW statistics such that $S \le 1 - 2\alpha$, for some significance level $\alpha \in (0,1)$, the null hypothesis is rejected after application of the DBN and uniformity of $\mathbf{P}$ is assumed as a consequence of cor. $(\ref{coraltthm})$. Finally, the existence of the critical $\epsilon_{c} > 0$ is illustrated by this example.  Given cors. $(\ref{coraltthm},\ref{coraltthm2})$ and thms. $(\ref{distthm},\ref{statthm},\ref{epsilonthm})$, this is the expected outcome.

\begin{table}[H]
\caption{KL \& SW Test Statistics by $\epsilon$}
\centering
\large
\begin{tabular}{c c c}
\hline\hline
$\epsilon$ & $E$ & $S$ \\ [0.5ex]
\hline
 70 & 63.1816 & 0.8541 \\
 80 & 53.2693 & 0.8574 \\
 90 & 45.0366 & 0.8634 \\
100 & 38.3761 & 0.8745 \\
110 & 32.2547 & 0.8913 \\
120 & 26.5077 & 0.9167 \\
130 & 20.9989 & 0.9431 \\
140 & 18.0090 & 0.9597 \\ [1ex]
\hline
\end{tabular}
\caption*{Table $(\ref{table:klswtab})$:  Decreasing entropy and increasing likelihood of normality of the distribution of intensities in the reconstruction for increasing $\epsilon > 0$ requires that the DBN applied to the reconstruction obtains the underlying equilibrium distribution of intensities from the original image, by thm. $(\ref{distthm},\ref{statthm})$ and uniqueness of $\mathbf{P}$ in $\cite[Thm.\ (8.1)]{Grimmett}$}
\label{table:klswtab}
\end{table}

\section{Conclusion}

It was shown that an edge detection heuristic could be devised to perform image segmentation and object localization of white-shifted pixels by estimating the distribution of intensities within an image and exploiting properties of its resulting distribution.  Indeed, it was shown that if an image is partitioned into sub-regions of a certain size and the edge detection heuristic is applied to a single sub-region within the image and extended to its boundary, then the resulting posterior distribution is maximal in the sense that image intensities across the whole image are increasingly correctly clustered with like values, allowing for a shift of intensity to highlight an object.  The shifted intensities increasingly mask the object in the original image so that the edge detection heuristic, supplied by a DBN that mimics an annealing process, offers an unsupervised method for object detection, with increasing accuracy as intensity differences within bounded sub-regions are allowed to grow larger.

\section{Proofs}

\subsection{Theorem\ $\ref{postthm}$}

\begin{proof}
Follows directly from the Kolmogorov Theorem and thm. $(\ref{weakthm})$. %$\qed$
\end{proof}

\subsection{Theorem\ $\ref{weakthm}$}

\begin{proof}
Given $\epsilon > 0$, fix $\delta > 0$ and define $\mathcal{C}^{\delta} = \{\mathbf{P}_{t \leftrightarrow t^{\prime}}^{\delta}:t^{\prime} \in \mathcal{B}_{t} \backslash \{t\}\}$ under the condition $\mathbf{P}_{t \leftrightarrow t^{\prime}}(|X_{t}-X_{t^{\prime}}| \ge \epsilon) = \delta$.  Then, the Kolmogorov Theorem can be applied to find a unique $\mathbf{P}^{\delta}$ whose single-edge conditionals are in $\mathcal{C}^{\delta}$, up to sets of $\mathbf{P}^{\delta}$-measure zero.  For $t \in \mathcal{B}$, let $\Omega_{t}$ be a subset of configurations of tuples which are fixed on $\mathcal{B} \backslash \mathcal{B}_{t}$ and let $\mathcal{A}_{t}$ be a $\sigma$-algebra of subsets of $\Omega_{t}$.  Clearly $X_{t}^{-1}(\mathcal{I}_{t}) \in \mathcal{A}_{t}$.  By the Kolmogorov Theorem, $\mathbf{P}_{t \leftrightarrow t^{\prime}}^{\delta}X_{t}^{-1}(\mathcal{I}_{t}) = \mathbf{P}^{\delta}X_{t}^{-1}(\mathcal{I}_{t})$ and $\mathbf{P}_{t \leftrightarrow t^{\prime}}X_{t}^{-1}(\mathcal{I}_{t}) = \mathbf{P}X_{t}^{-1}(\mathcal{I}_{t})$.  Therefore,
\begin{equation}
\label{weakeq1}
\begin{split}
\mathbf{P}^{\delta}X_{t}^{-1}(\mathcal{I}_{t})-\mathbf{P}X_{t}^{-1}(\mathcal{I}_{t}) &= \bigg(\mathbf{P}^{\delta}X_{t}^{-1}(\mathcal{I}_{t})-\mathbf{P}_{t \leftrightarrow t^{\prime}}^{\delta}X_{t}^{-1}(\mathcal{I}_{t})\bigg)\\&+\bigg(\mathbf{P}_{t \leftrightarrow t^{\prime}}^{\delta}X_{t}^{-1}(\mathcal{I}_{t})-\mathbf{P}_{t \leftrightarrow t^{\prime}}X_{t}^{-1}(\mathcal{I}_{t})\bigg)\\&+\bigg(\mathbf{P}_{t \leftrightarrow t^{\prime}}X_{t}^{-1}(\mathcal{I}_{t})-\mathbf{P}X_{t}^{-1}(\mathcal{I}_{t})\bigg)\\&=\mathbf{P}_{t \leftrightarrow t^{\prime}}^{\delta}X_{t}^{-1}(\mathcal{I}_{t})-\mathbf{P}_{t \leftrightarrow t^{\prime}}X_{t}^{-1}(\mathcal{I}_{t}).
\end{split}
\end{equation}
Allowing $\delta \rightarrow 0$ in eq. $(\ref{weakeq1})$, we have $\mathbf{P}_{t \leftrightarrow t^{\prime}}^{\delta}X_{t}^{-1}(\mathcal{I}_{t})-\mathbf{P}_{t \leftrightarrow t^{\prime}}X_{t}^{-1}(\mathcal{I}_{t}) \rightarrow 0$ by the definition of $\mathbf{P}_{t \leftrightarrow t^{\prime}}^{\delta}$.  Thus, $\mathbf{P}^{\delta}X_{t}^{-1}(\mathcal{I}_{t})-\mathbf{P}X_{t}^{-1}(\mathcal{I}_{t}) \rightarrow 0$, which implies $\mathbf{P}^{\delta} \xrightarrow{W} \mathbf{P}$ by def. $(\ref{weakdef})$. %$\qed$
\end{proof}

\subsection{Lemma\ $\ref{symlem}$}

\begin{proof}
Seeking a contradiction, suppose no such $\delta > 0$ exists.  Then, $\mathbf{P}(|X_{t}-X_{t^{\prime}}| \ge \epsilon) = 0$ when $|i_{t}-i_{t^{\prime}}| \ge \epsilon > 0$ for some chosen $\epsilon > 0$.  Yet, the edge $e_{t \rightarrow t^{\prime}} \in \mathbf{E}$ is open under the condition $\mathbf{P}(|X_{t}-X_{t^{\prime}}| \ge \epsilon) = 0$, requiring $|i_{t}-i_{t^{\prime}}| = 0$.  %$\qed$
\end{proof}

\subsection{Lemma\ $\ref{updlem}$}

\begin{proof}
If $\mathbf{P}(|i_{t}-i_{t^{\prime}}| = 0) = 1$ after successive updates are applied to an edge $e_{t \leftrightarrow t^{\prime}} \in \mathbf{E}$ for $t,t^{\prime} \in \mathcal{B}$ such that $t^{\prime} \in \mathcal{B}_{t}^{m} \backslash \{t\}$ and $t \in \mathcal{B}_{t^{\prime}}^{m} \backslash \{t^{\prime}\}$, then by independence of updates applied in each overlapping region, the point process $\mathcal{P}$ is sampling from the equilibrium distribution.  Thus, without loss of generality, assume the open edge probability, $\mathbf{P}(|i_{t}-i_{t^{\prime}}| = 0) = 1-q$ for some $q \in (0,1)$.  Then, $\mathbf{P}(|i_{t}-i_{t^{\prime}}| > 0) = q \in (0,1)$.  %$\qed$
\end{proof}

\subsection{Corollary\ $\ref{stochcor}$}

\begin{proof}
Follows directly from lem. $(\ref{symlem})$ and thms. $(\ref{postthm},\ref{weakthm})$.  %$\qed$
\end{proof}

\subsection{Proposition\ $\ref{critprop}$}

\begin{proof}
Follows directly from statements preceding the statement of prop. $(\ref{critprop})$.  %$\qed$
\end{proof}

\subsection{Theorem\ $\ref{critmthm}$}

\begin{proof}
Follows directly from $\cite{Murphy}$ and statements preceding the statement of thm. $(\ref{critmthm})$ that
\begin{equation}
\begin{split}
m_{c} = \floor[\bigg]{\sqrt{\frac{d}{K_{c}^{2}}}} = \floor[\bigg]{\sqrt{\frac{m^{2}}{K_{c}^{2}}}} = \floor[\bigg]{\frac{m}{K_{c}}}.
\end{split}
\end{equation}
%$\qed$
\end{proof}

\subsection{Theorem\ $\ref{Rdecaythm}$}

\begin{proof}
Follows directly from statements following the statement of thm. $(\ref{Rdecaythm})$ and from an application of $\cite[Thm.\ (6.30)]{Grimmett2}$.  %$\qed$
\end{proof}

\subsection{Theorem\ $\ref{altthm}$}

\begin{proof}
Follows directly from statements preceding and following the statement of thm. $(\ref{altthm})$.  %$\qed$
\end{proof}

\subsection{Corollary\ $\ref{coraltthm}$}

\begin{proof}
The case $|\mathbf{S}| = 2$ follows directly from thm. $(\ref{altthm})$ and the statements between the statement of thm. $(\ref{altthm})$ and the statement of cor. $(\ref{coraltthm})$.  Suppose $|\mathbf{S}| = n > 2$.  In the induction step, for $|\mathbf{S}| = n-1 \ge 2$, the equilibrium distribution is uniform by assumption.  By applying thm. $(\ref{altthm})$ again, we can find a conditional equilibrium distribution for any $2$ arbitrary, bounded sub-regions, where one of the bounded sub-regions almost surely contains only edges having the missing state.  After annealing, its posterior is uniform by the case $|\mathbf{S}| = 2$.  The corollary follows by the Kolmogorov Theorem.  %$\qed$
\end{proof}

\subsection{Corollary\ $\ref{coraltthm2}$}

\begin{proof}
Follows directly from an application of the Central Limit Theorem $\cite{Guyon,Hogg}$, along with cor. $(\ref{stochcor})$ and prop. $(\ref{critprop})$.  %$\qed$
\end{proof}

\subsection{Theorem\ $\ref{distthm}$}

\begin{proof}
Follows directly from statements preceding the statement of thm. $(\ref{distthm})$.  %$\qed$
\end{proof}

\subsection{Corollary\ $\ref{distcor}$}

\begin{proof}
Follows directly from thm. $(\ref{distthm})$ and cor. $(\ref{coraltthm2})$.  %$\qed$
\end{proof}

\subsection{Theorem\ $\ref{statthm}$}

\begin{proof}
Follows directly from statements preceding the statement of thm. $(\ref{statthm})$.  %$\qed$
\end{proof}

\subsection{Theorem\ $\ref{epsilonthm}$}

\begin{proof}
Follows directly from statements following the statement of thm. $(\ref{epsilonthm})$.  %$\qed$
\end{proof}

\end{document}